\documentstyle[aps,prd,amsfonts,amssymb,twocolumn]{revtex}
\input epsf

\newcommand{\slL}{\raise.15ex\hbox{$/$}\kern-.53em\hbox{$L$}}
\newcommand{\slP}{\raise.15ex\hbox{$/$}\kern-.53em\hbox{$P$}}
\newcommand{\slR}{\raise.15ex\hbox{$/$}\kern-.53em\hbox{$R$}}
\newcommand{\slQ}{\raise.15ex\hbox{$/$}\kern-.53em\hbox{$Q$}}
\newcommand{\slK}{\raise.15ex\hbox{$/$}\kern-.53em\hbox{$K$}}
\newcommand{\slSigma}{\raise.15ex\hbox{$/$}\kern-.53em\hbox{$\Sigma$}}
\newcommand{\slcalP}{\raise.15ex\hbox{$/$}\kern-.63em\hbox{$\cal P$}}
\newcommand{\be}{\begin{equation}}
\def\build#1\over#2{\mathrel{\mathop{\kern 0pt#1}\limits_{#2}}}

\begin{document}
\title{\bf{Quasi-real photon production\\
    in thermal QCD\thanks{Work done in collaboration with P.~Aurenche,
      R.~Kobes, E.~Petigirard and H.~Zaraket. Talk presented at the
      TFT'98 conference, Regensburg, Germany, 10-14 august 1998.}}}
\author{F.~Gelis}
\address{Laboratoire de Physique Th\'eorique LAPTH,\\
  BP110, F-74941, Annecy le Vieux Cedex, France} \date{\today}

\maketitle

\begin{abstract}
  In this proceedings, I consider two-loop contributions to real
  photon production in thermal QCD. If the photon is strictly
  massless, strong collinear divergences appear in this calculation.
  These singularities are regularized by the quark thermal mass of
  order $gT$, which generates powers of $1/g$, so that the
  corresponding terms are strongly enhanced with respect to naive
  expectations of their order of magnitude.
\end{abstract}
\vskip 5mm
LAPTH-98/699, hep-ph/9809380

\section{Introduction}
From a phenomenological point of view, photon production is considered
as a potential candidate for the detection and discrimination of a
quark gluon plasma. This quantity is also interesting from a more
formal point of view in order to test our control over some aspects of
thermal field theories like infrared and collinear singularities. One
reason is that this quantity is obviously observable and should
therefore be infrared and collinear safe in a consistent theory.

At one loop, the thermal production of soft real photons
\cite{BaierPS1,AurenBP1} suffers from a logarithmic collinear
singularity, that can be cured by an extension of the hard thermal
loop resummation technique proposed by Flechsig and Rebhan
\cite{FlechR1}. It appeared also that common processes like
bremsstrahlung are not present at one loop, while they are assumed to
play the dominant role in semi-classical approaches
\cite{CleymGR2,CleymGR3,BaierDMPS1,BaierDMPS2}.

Therefore, we considered the two-loop diagrams giving bremsstrahlung
\cite{AurenGKP1,AurenGKP2,AurenGKZ1}.  It turns out that they suffer
from collinear singularities much stronger than the 1-loop ones. After
regularization by a thermal mass of order $gT$, these singularities
give extra factors $1/g$, which spoils the natural hierarchy of
perturbative expansion. I present this calculation in this
proceedings. Then, I compare the result with that of other approaches,
and finally, I conclude with some open questions related to what might
happen at higher orders.

\section{Tools}
\subsection{Hard thermal loops}
This first section is devoted to a presentation of the tools we used
in order to perform this calculation. The general framework of our
study is the effective perturbative expansion that one obtains after
the resummation of hard thermal loops \cite{BraatP1,FrenkT1}. To be
more definite, we used the retarded-advanced version of the real time
formalism \cite{AurenB1,EijckKW1}. Initially, one of the motivations
for hard thermal loops was the resolution of the long standing puzzle
of the gluon damping rate, but they provided also a consistent
framework to deal with the infrared sector of thermal theories.

\subsection{Counterterms}
A possible way to implement calculations in this effective theory is
to use cut-offs \cite{BraatT1}: loops with effective propagators and
vertices are evaluated with an upper cut-off $\Lambda$ ($gT\ll \Lambda
\ll T$).  If higher order corrections are not necessary, the cut-off
dependence is usually of a smaller order of magnitude. If on the
contrary an important dependence on $\Lambda$ is observed, then one
must include higher order diagrams correcting this loop, in which
$\Lambda$ must be used as a lower cut-off to prevent double counting.
\begin{figure}[htbp]
  \centerline{
    \mbox{\epsfxsize=8cm\epsffile{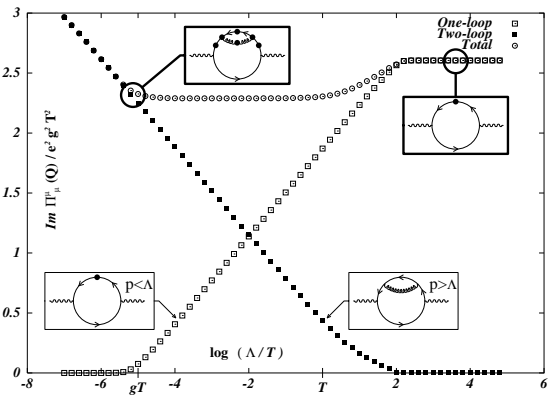}}}
  \caption{\footnotesize{Cancellation of cut-off 
      dependence illustrated on the example of Baier et al. This
      figure also shows the 1-loop and 2-loop contributions when one
      does not use a cut-off (diagrams in boldface boxes).}}
  \label{fig:cut-off}
\end{figure}
One should then check that the sum of both terms is $\Lambda$
independent at the desired order. This procedure has been used also in
\cite{BaierNNR1} to calculate the hard real photon production rate.
The results are illustrated graphically on figure \ref{fig:cut-off},
which represents numerical evaluations of one- and two-loop
contributions to the photon polarization tensor.

This procedure can also be interpreted more formally in terms of an
effective lagrangian and counterterms.  The proper way to perform the
hard thermal loop resummation is to write an effective lagrangian
${\cal L}_{\rm eff}$ which is the sum of the bare lagrangian $\cal L$
and of the term ${\cal L}_{_{\rm HTL}}$ containing all the hard
thermal corrections \cite{BraatP4,FrenkT2}. Therefore, it appears
clearly that one needs to subtract counterterms in order to keep the
overall lagrangian unmodified. Basically, this procedure amounts to
write
\begin{eqnarray}
&&{\cal L}={\cal L}_{\rm eff}+{\cal L}_{\rm ct}\\
&&{\cal L}_{\rm eff}={\cal L}+{\cal L}_{_{\rm HTL}}\\
&&{\cal L}_{\rm ct}=-{\cal L}_{_{\rm HTL}}\; .
\end{eqnarray}
That way, the HTL resummation is just a mere reordering of the
perturbative expansion, and the exact theory remains the same.
Counterterms have been used in \cite{Schul1} for the two-loop plasmon
mass.

The subtraction of counterterms and the use of cut-offs are equivalent.
This is illustrated on figure \ref{fig:cut-off}, where we also
represented the 1-loop and 2-loop contributions when the cut-off is
removed, in the boldface boxes. Obviously, their sum is much higher
than the value of the plateau in the cut-off approach. This means that
this sum includes some contributions twice, and that a counterterm is
mandatory.

Counterterms seem more systematic since they come from a lagrangian
formulation, while it can be tricky to use properly cut-offs in a
situation with overlapping loops. In the following, I follow the point
of view based on counterterms.

\subsection{Photon production rate}
Let me now give the precise relationship between the photon production
rate and the thermal Green's functions. The number of real photons
emitted per unit time and per unit volume of the plasma is related to
the retarded photon self energy via \cite{Weldo3,GaleK1}
\begin{equation}
\label{eq:rate}
{{dN}\over{dtd^3\bbox{x}}}=-{{d^3\bbox{q}}\over{(2\pi)^32q_o}}
2n_{_{B}}(q_o) {\rm Im}\,\Pi^{^{RA}}{}_\mu{}^\mu(q_o,\bbox{q})\; ,
\end{equation}
where $n_{_{B}}(q_o)\equiv 1/(\exp(q_o/T)-1)$ is the Bose-Einstein
statistical factor. For photons of invariant mass $Q^2>0$ subsequently
decaying in a lepton pair, we have instead
\begin{equation}
{{dN}\over{dtd^3\bbox{x}}}=-{{dq_od^3\bbox{q}}\over{12\pi^4}}
{\alpha\over{Q^2}}\,
n_{_{B}}(q_o) {\rm Im}\,\Pi^{^{RA}}{}_\mu{}^\mu(q_o,\bbox{q})\; .
\end{equation}
Basically, the second formula differs from the first one by the allowed
phase space, by the propagator $1/Q^2$ of a massive photon, and by the
coupling constant $\alpha$ which appears in the decay into a lepton
pair\footnote{These formulas are valid only at first order in the QED
  constant $\alpha$ since they neglect any reinteraction of the
  emitted photon on its way out of the plasma. On the contrary, they
  include corrections to all orders in the strong coupling constant.}.
The imaginary part of the retarded self energy appearing in the right
hand side can be expressed as a sum of cuts through the corresponding
diagrams \cite{KobesS1,KobesS2,Gelis3}.

\section{Two-loop contributions}

\subsection{Involved topologies}

Bremsstrahlung appears in the two-loop diagrams represented on figure
\ref{fig:two-loop}.
{\begin{figure}[htbp]
  \centerline{
    \mbox{\epsfxsize=7cm\epsffile{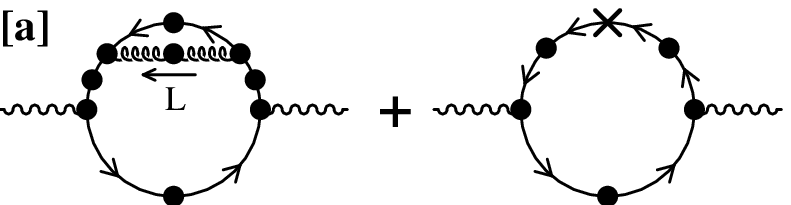}}
    }
  \vskip 2mm
  \centerline{
    \mbox{\epsfxsize=7cm\epsffile{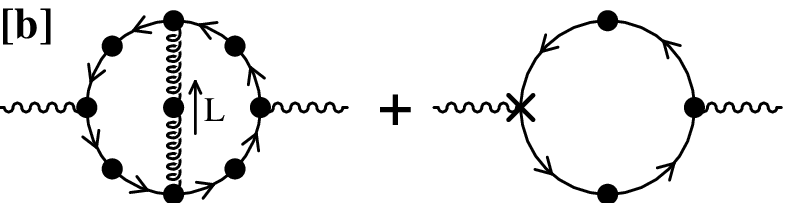}}
    }
\vskip 2mm
  \caption{\footnotesize{Two-loop contributions involving 
      bremsstrahlung processes. A black dot denotes an effective
      propagator or vertex. Crosses are HTL counterterms.}}
  \label{fig:two-loop}
\end{figure}}

{ According to the discussion made earlier, they should be accompanied
  by the 1-loop diagrams involving the proper HTL counterterms. In
  fact, these two-loop diagrams contain much more than bremsstrahlung.
  For instance, if we consider a cut that goes through the gluon
  propagator, we can have bremsstrahlung if the gluon is space-like
  ($L^2<0$) and corrections to Compton effect if $L^2>0$.  Concerning
  the counterterms, since their role is to avoid double counting, it
  is quite natural to associate them with the contributions coming
  from the $L^2>0$ region of phase space. Indeed, the inner structure
  of the counterterm involves a bare gluon which contributes only in
  the time-like region once cut. Another way to justify this is to say
  that the counterterms should be associated with terms that provide
  corrections to a process appearing already at a lower order. There
  is no point in associating counterterms with a process that appears
  in this order for the first time.  Therefore, counterterms will be
  calculated only if it appears that the $L^2>0$ region contributes at
  leading order.  }

\begin{figure}[htbp]
  \centerline{ \mbox{\epsfxsize=7cm\epsffile{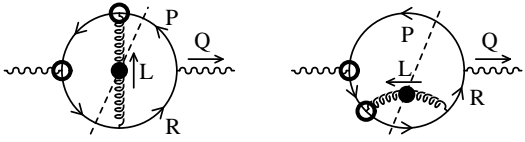}}}
  \caption{\footnotesize{Simplified two-loop contributions. The circled 
      vertices correspond to the framework of the cutting rules.}}
  \label{fig:two-loop-simple}
\end{figure}

Phase space considerations tend to make us favor a hard quark loop in
order to enlarge its size as much as possible.  Therefore, we can drop
at leading order hard thermal corrections involving the propagator of
such a hard quark, and obtain a simplified version of the previous
diagrams, which is represented on figure \ref{fig:two-loop-simple}.
More precisely, we will use the hard limit of the effective quark
propagator, retaining that way an asymptotic thermal mass.  This mass
will be necessary later in order to regularize collinear
singularities.\footnote{Contrary to \cite{FlechR1}, we don't have to
  introduce such a mass by hand in the theory, since it comes
  naturally from the diagrams we are considering.} Other authors
\cite{Niega5,Niega6} have found that the width obtained by an extra
resummation including a self energy with an imaginary part in the time
like region may compete with this asymptotic mass as a collinear
regulator under certain circumstances. But since our goal was to
follow strictly the HTL approach, we did not consider this extension
here.

\subsection{Matrix element}
We have checked that the two cuts represented on figure
\ref{fig:two-loop-simple} form a gauge invariant set of terms. One
should also add a second symmetrical cut to the vertex, and a second
self-energy diagram in which the loop correction is on the other quark
line. We can as well take them into account by just multiplying the
result by an extra factor $2$, since they provide contributions equal
to those of the cuts already represented.  Applying the cutting rules
for the retarded-advanced formalism \cite{Gelis3} gives immediately
for the vertex contribution
\begin{eqnarray}
&&
  {\rm Im}\,\Pi^{^{RA}} {}_\mu{}^\mu(q_o,{\bbox{q}})_{|_{\rm
      vertex}}\nonumber\\
  &&=-{{NC_{_{F}}}\over 2}e^2g^2
  \int{{d^4P}\over{(2\pi)^4}}\!\!\!\int{{d^4L}\over{(2\pi)^4}}
  \left[\Delta^{^{R}}_{_{T,L}}(L)\!-\!\Delta^{^{A}}_{_{T,L}}(L)\right]
  \nonumber\\
  &&\times\;
  \left[S^{^{R}}(P)-S^{^{A}}(P)\right]
  \left[S^{^{R}}(R+L)-S^{^{A}}(R+L)\right]
  \nonumber\\
  &&\times\;
  \left(n_{_{F}}(r_o)-n_{_{F}}(p_o)\right)
  \left(n_{_{B}}(l_o)+n_{_{F}}(r_o+l_o)\right)
  \nonumber\\
  &&\times\;
  S(R)S(P+L)P_{\rho\sigma}^{^{T,L}}(L)\;
  {\rm Trace}^{\rho\sigma}{}_{|_{\rm vertex}}\; ,
\end{eqnarray}
where, we denote the fermion propagator:
\begin{eqnarray}
  &&{\cal S}^{^{R,A}}(P)\equiv{\overline{\slP}} S^{^{R,A}}(P){\ \rm with\ }
  {\overline{P}}\equiv(p_o,\sqrt{p^2+M^2_\infty}\,\hat{\bbox{p}})\\
  &&S(P)^{^{R,A}}
  \equiv{i\over{{\overline{P}}^2\pm ip_o\varepsilon}}
  ={i\over{P^2-M^2_{\infty}\pm ip_o\varepsilon}}\; ,
  \label{notations:quark}
  \end{eqnarray}
  and the effective gluon propagator in linear covariant gauges:
 \begin{eqnarray}
  &&-D_{\rho\sigma}^{^{R,A}}(L)\nonumber\\
&&\;\equiv
  P^{^{T}}_{\rho\sigma}(L)\Delta^{^{R,A}}_{_{T}}(L)
  +P^{^{L}}_{\rho\sigma}(L)\Delta^{^{R,A}}_{_{L}}(L)
  +\xi L_\rho L_\sigma/L^4\\
  &&\Delta^{^{R,A}}_{_{T,L}}(L)\equiv
  \left.{i\over{L^2-\Pi_{_{T,L}}(L)}}\right|_{_{R,A}}\nonumber\\
  &&\Pi_{_{T}}(L)\equiv3m_{\rm g}^2\left[
    {{x^2}\over 2}+{{x(1-x^2)}\over{4}}\ln\left(
      {{x+1}\over{x-1}}\right)\right]\\
  &&\Pi_{_{L}}(L)\equiv3m_{\rm g}^2(1-x^2)\left[
    1-{{x}\over{2}}\ln\left(
      {{x+1}\over{x-1}}\right)\right]\; ,
  \label{notations:gluon}
\end{eqnarray}
\noindent
with $P^{^{T,L}}_{\rho\sigma}$ the usual transverse and longitudinal
projectors \cite{Pisar6,Weldo1,LandsW1}, $M^2_{\infty}\equiv g^2
C_{_{F}} T^2/4$ \cite{FlechR1} the asymptotic thermal mass of the
quark, $m_{\rm g}^2 \equiv g^2T^2[N+N_{_{F}}/2]/9$ the soft gluon
thermal mass, and $x\equiv l_o/l$.  In this formula, $e$ denote the
electric charge of the quark and therefore depends on its flavor.
Likewise, we obtain for the second diagram:
\begin{eqnarray}
  &&
  {\rm Im}\,\Pi^{^{RA}}
  {}_\mu{}^\mu(q_o,{\bbox{q}})_{|_{\rm self}}\nonumber\\
&&=-{{NC_{_{F}}}\over 2}e^2g^2
  \int{{d^4P}\over{(2\pi)^4}}\int{{d^4L}\over{(2\pi)^4}}
  \left[\Delta^{^{R}}_{_{T,L}}(L)-\Delta^{^{A}}_{_{T,L}}(L)\right]
  \nonumber\\
  &&\times\;
  \left[S^{^{R}}(P)-S^{^{A}}(P)\right]
  \left[S^{^{R}}(R+L)-S^{^{A}}(R+L)\right]
  \nonumber\\
  &&\times\;
  \left(n_{_{F}}(r_o)-n_{_{F}}(p_o)\right)
  \left(n_{_{B}}(l_o)+n_{_{F}}(r_o+l_o)\right)\nonumber\\
&&\times\;
\left(S(R)\right)^2
P_{\rho\sigma}^{^{T,L}}(L)\;
  {\rm Trace}^{\rho\sigma}{}_{|_{\rm self}}\; .
  \label{cutself}
\end{eqnarray}
For propagators $S(R)$ without any $R$ or $A$ superscript, this
prescription is irrelevant because of the position of the cuts. It
means that only the principal part of these propagators contributes.

Taking into account the identity $L^\rho
P^{^{T,L}}_{\rho\sigma}(L)=0$, as well as the fact that certain terms
will be killed later by the $\delta(.)$ distributions, we can give the
following expressions for the Dirac's traces
\begin{eqnarray}
  &&{\rm Trace}{}^{\rho\sigma}{}_{|_{\rm self}}
  \approx
  -4\left[ 4{\overline{R}}{}^2Q{}^\rho {\overline{R}}{}^\sigma 
    -4Q{}^2{\overline{R}}{}^\rho {\overline{R}}{}^\sigma\right.
  \nonumber\\
  &&\quad\left.-g{}^{\rho\sigma}\left(
      {\overline{R}}{}^2({\overline{R}}{}^2-Q{}^2)
      +2{\overline{R}}{}^2Q\cdot L-2Q{}^2
      {\overline{R}}\cdot L\right)\right]\; ,\\
  &&{\rm Trace}{}^{\rho\sigma}{}_{|_{\rm vertex}}
  \approx
  -4\left[2{\overline{R}}{}^2{\overline{P}}{}^\rho Q{}^\sigma
    -2({\overline{P+L}}){}^2{\overline{R}}{}^\rho Q{}^\sigma
  \right.
  \nonumber\\
  &&
  \quad
  +2L{}^2({\overline{R}}{}^\rho {\overline{R}}{}^\sigma
  +{\overline{P}}{}^\rho {\overline{P}}{}^\sigma)
  -4Q{}^2{\overline{R}}{}^\rho {\overline{P}}{}^\sigma
  \nonumber\\
  &&\quad+g{}^{\rho\sigma}\left(
    \left.-L{}^2({\overline{R}}{}^2+({\overline{P+L}}){}^2-Q{}^2-L{}^2)
    \right)\right]\; .
\end{eqnarray}

\subsection{Phase space}
\begin{figure}[ht]
  \centerline{\epsfxsize=7cm\epsffile{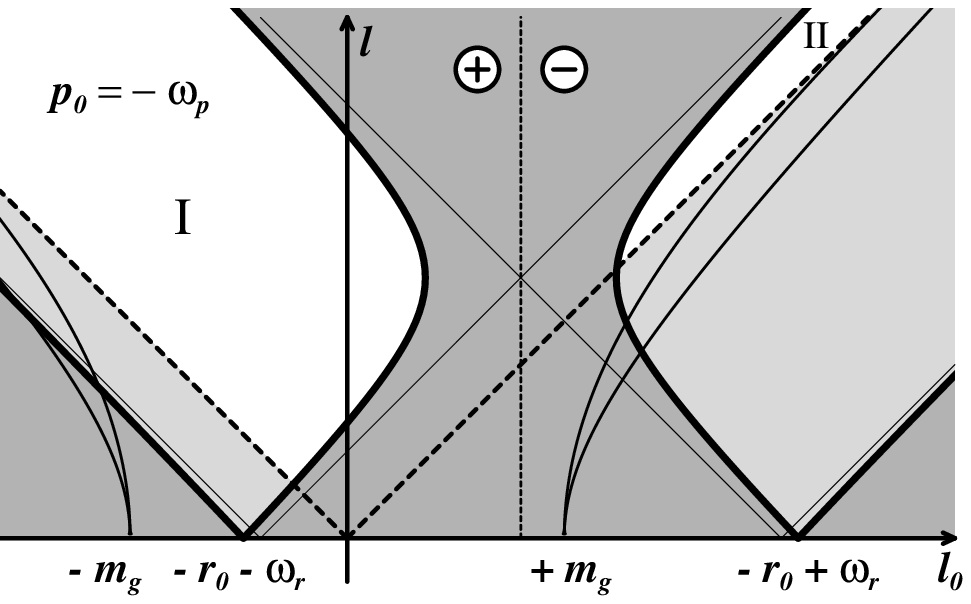}}
  \centerline{\epsfxsize=7cm\epsffile{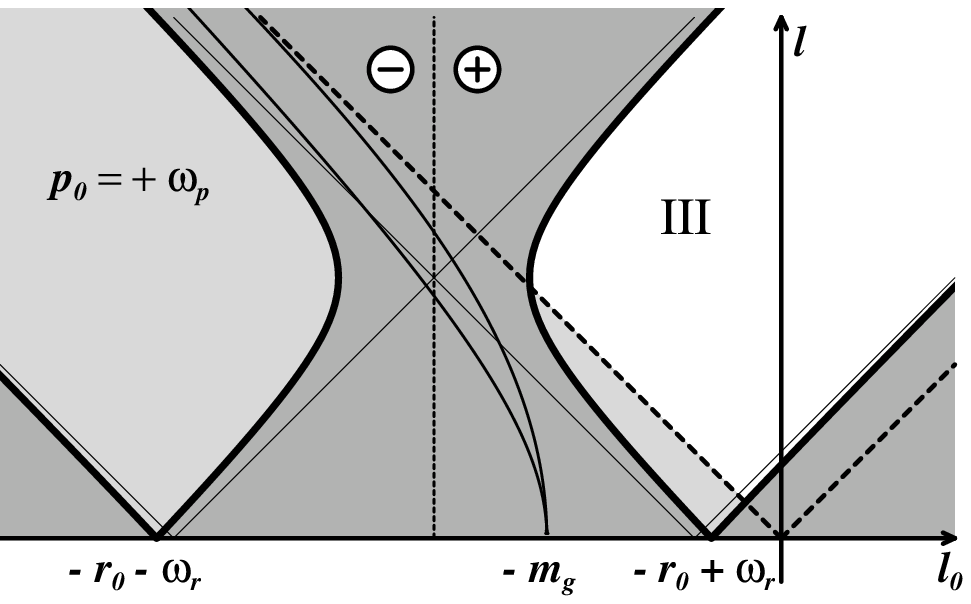}}
  \caption{\footnotesize{Allowed domains in the $(l_o,l)$ plane for 
      $p_o=\pm \omega_p$. 
      The area shaded in dark
      gray is excluded by the delta function constraints.
      The areas shaded in light gray are above the light-cone (dotted lines).
      The light curves are the transverse and
      longitudinal mass shells of the thermalized gluon. 
      The vertical dotted
      line is the separation between $\epsilon(p_o)\epsilon(r_o+l_o)=+1$ and 
      $\epsilon(p_o)\epsilon(r_o+l_o)=-1$.}}
  \label{fig:llo-reduc}
\end{figure}

\begin{figure}[ht]
  \centerline{\epsfxsize=7cm\epsffile{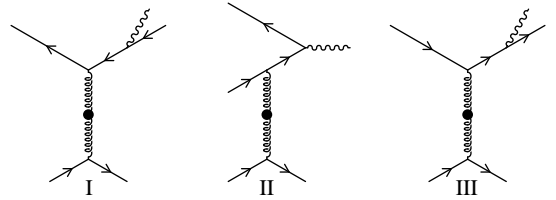}}
  \caption{\footnotesize{Physical processes included in the 
      regions where $L^2<0$.}}
  \label{fig:processes-list}
\end{figure}

Let me now study the restrictions that the $\delta(.)$ functions
impose on the available phase space. Since we may encounter collinear
divergences, we need to keep carefully the asymptotic mass $M_\infty$
in this study. From the first cut propagator
$S^{^{R}}(P)-S^{^{A}}(P)=2\pi\epsilon(p_o)\delta(P^2-M^2_\infty)$, we
obtain $p_o=\pm\omega_{\bbox{p}}$, where I denote
$\omega_{\bbox{p}}\equiv\surd({\bbox{p}}^2+M^2_\infty)$, as well as
$r_o=p_o+q_o$. The second delta function constrains the angle
$\theta'$ between the 3-vectors $\bbox{r}$ and $\bbox{l}$, by
\begin{equation}
\cos\theta^\prime={{R^2-M^2_\infty+2r_ol_o+L^2}\over{2rl}}\; .
\end{equation}
At that point, we must impose that this cosine be between $-1$ and
$+1$, which amounts to the following set of inequalities
\begin{eqnarray}
  &&(r_o-r+l_o+l)(r_o+r+l_o-l)\ge M^2_\infty\nonumber\\
  &&(r_o-r+l_o-l)(r_o+r+l_o+l)\le M^2_\infty\; .
  \label{hardconstraints}
\end{eqnarray}
A convenient way is to see these inequalities as constraints in the
$(l_o,l)$ plane in which they reduce the integration domain. The
effect of this reduction is represented on figure \ref{fig:llo-reduc}.
On this figure, regions shaded in dark gray are excluded by the
inequalities \ref{hardconstraints}. Regions shaded in light gray are
allowed and involve a time-like cut gluon propagator, which means
that the corresponding processes correspond to Compton or annihilation
processes. Regions in white are also allowed and correspond to the
exchange of a space-like gluon. These regions will turn out to be
dominant.
The regions on which we focus mostly in the following have been
labeled by I, II and III. Looking at the signs of the quark momenta in
each region, we can see that region I and III involve bremsstrahlung
processes, while a new process appears in region II.  This unexpected
process can be described as a $q\bar{q}$ annihilation in which one of
the fermions undergoes a scattering through a gluon exchange (see the
figure \ref{fig:processes-list}).
By the change of variables $P\to -R-L$, we can map the region I on the
region III. As a consequence, the bremsstrahlung photon production can
be obtained by looking only at the region III, and multiplying its
contribution by an overall factor $2$.

\subsection{Collinear singularities}
When both the external photon and the internal quark are massless, the
diagrams of figure \ref{fig:two-loop-simple} display a collinear
singularity when the 3-momentum of the quark is parallel to that of
the emitted photon. Moreover, contrary to the collinear divergences
encountered in the 1-loop contributions, which are at most
logarithmic, we have here two denominators that can vanish almost
simultaneously. This is of course obvious for the second diagram of
figure \ref{fig:two-loop-simple} in which we have a double pole. But
even for the first one, the poles are distinct but very close. Indeed,
$R^2$ is vanishing when $\bbox{p}$ is parallel to $\bbox{q}$, while
$(P+L)^2$ vanishes when $\bbox{r}+\bbox{l}$ is parallel to $\bbox{q}$.
As a consequence, when $\bbox l$ is soft, these two conditions of
collinearity are satisfied almost simultaneously, independently of the
order of magnitude of the momentum $\bbox{q}$. Therefore, the two
denominators have very close poles, and their association mostly
behaves like a double pole.

The collinear singularity of these diagrams is therefore quasi linear,
instead of logarithmic.  When the quark mass is taken into account,
these divergences are regularized, but a consequence of their linear
nature is the generation of powers of $1/g$. The strength of these
divergences makes dominant the terms that have this configuration of
quasi double poles. It is rather straightforward to show that this
cannot occur when the gluon is time-like. Indeed, because of the delta
function $\delta(P^2-M^2_\infty)$, the denominator
$(P+L)^2-M^2_\infty$ can approach zero only if $L^2<0$ for soft
$L$.\footnote{The configuration with $L^2=0$ and hard $L$ has in fact
  been studied by \cite{BaierNNR1,KapusLS1} (there is no contribution
  with $L^2>0$ in \cite{BaierNNR1,KapusLS1} since they used bare gluon
  propagators).} Therefore, we consider only the regions where
$L^2<0$.  It is easy to check that in the collinear limit, for soft
$L$, the boundaries of region III are just $-l\le l_o\le l$, {\it
  i.e.\/} the inequality $|\cos\theta'|\le 1$ does not add any extra
constraint beyond $L^2 < 0$.

\subsection{Soft photon production}
Let me start with soft photons. I consider here photons having a small
invariant mass $Q^2\ll q_o^2\ll T^2$.\footnote{The previous discussion
  on collinear singularities still holds for photons which are
  slightly virtual. The inequality $Q^2\ll q_o^2$ is in fact the
  condition on the invariant mass in order to have this collinear
  enhancement.} In this situation, it is easy to verify that the
contribution of region II is subdominant, since the area of the phase
space available to this process in the $(l_o,l)$ plane is proportional
to $q_o$. Moreover, for this region, the gluon momentum $L$ is
necessarily hard, which implies that the two conditions of
collinearity discussed above cannot be satisfied simultaneously.
Therefore, bremsstrahlung is the only dominant process for the
production of low invariant mass soft photons.

Using only the variables $u=1-\cos\theta$ where $\theta$ is the angle
between $\bbox{p}$ and $\bbox{q}$, $p$, $l$ and $x=l_o/l$, we can
write\cite{AurenGKP2}
\begin{equation}
  R^2-M^2_\infty\approx 2pq\left[u+{{M^2_{\rm eff}}\over{2p^2}}\right]\; ,
  \label{eq:R2}
\end{equation}
where $M_{\rm eff}^2\equiv M^2_\infty+Q^2p^2/q_o^2$, and
\begin{equation}
  \int\limits_{0}^{2\pi}\!\!\!{{d\phi}\over{(P+L)^2-M^2_\infty}}\approx
  {{2\pi(2pq)^{-1}}\over{\left[
        \left(u+{{M^2_{\rm eff}}\over{2p^2}}+{{L^2}\over{2p^2}}\right)^2
        -{{L^2}\over{p^2}}{{M^2_{\rm eff}}\over{p^2}}\right]^{1/2}}}\; ,
\label{eq:PL2}
\end{equation}
where we performed the integration over the azimuthal angle $\phi$
between $\bbox{q}$ and $\bbox{l}$ at this stage since this denominator
is the only place where $\phi$ appears at the dominant order. Among
all the terms present in the amplitude, the only ones to be sensitive
to the collinear singularity described in the previous section are
those where two of these denominators are present. Other terms have a
numerator that cancels one of the denominators, so that the collinear
singularity is only logarithmic.  On the basis of these
considerations, it is easy to check that only the following term in
$L^2$
\begin{equation}
 -8L^2{{\overline{R}{}^\rho\overline{R}{}^\sigma+
      \overline{P}{}^\rho\overline{P}{}^\sigma}
    \over{\overline{R}{}^2(\overline{P+L}){}^2}}\; 
\label{eq:dominant}
\end{equation}
is dominant.

It is then straightforward to perform the angular integration over
$u=1$, to obtain the following expression of the imaginary part of the
polarization tensor of the photon:
\begin{equation}
\hbox{\rm Im}\,\Pi^{^{RA}}{}_\mu{}^\mu(q_o,{\bbox{q}})\approx
  -\frac{e^2 g^2 N_{_{C}} C_{_{F}}}{3\pi^2}
  \frac{T^3}{q_0}(J_{_{T}}+J_{_{L}})
\label{eq:im-pi}
\end{equation}
where we denoted 
\begin{eqnarray}
\!\!\!J_{_{T,L}}\equiv&&{6\over{\pi^2}}\int_{0}^{+\infty}dv v^2\,
{e^v\over{(e^v+1)^2}}
\int\limits_{0}^{1}\;
\frac{dx}{x}\;|\widetilde{I}_{_{T,L}}(x)| \nonumber\\
&&\times\int\limits_{0}^{+\infty}\;dw\;
\frac{\sqrt{w/(w+4)}
\hbox{\rm tanh}{}^{-1}\sqrt{w/(w+4)}}{(w+\widetilde{R}_{_{T,L}}(x))^2+
(\widetilde{I}_{_{T,L}}(x))^2}\; ,
\label{imag}
\end{eqnarray}
with
\begin{eqnarray}
&&v\equiv {p\over T}\;,\;\quad w\equiv {{-L^2}\over{M^2_{\rm eff}}}\nonumber\\
&&\widetilde{I}_{_{T,L}}(x)\equiv {{\hbox{\rm Im}\,
\Pi_{_{T,L}}(x)}\over{M^2_{\rm eff}}}\;,\;\quad
\widetilde{R}_{_{T,L}}(x)\equiv {{\hbox{\rm Re}\,
\Pi_{_{T,L}}(x)}\over{M^2_{\rm eff}}}\; .
\end{eqnarray}

A first noticeable feature of the result of Eq. (\ref{eq:im-pi}) is
its order of magnitude. Indeed, it is $1/g^2$ larger than 1-loop
contributions calculated in \cite{BaierPS1,AurenBP1}. This is due to
the strong collinear singularities described earlier.

\begin{figure}[htbp]
  \centerline{\epsfxsize=7cm\epsffile{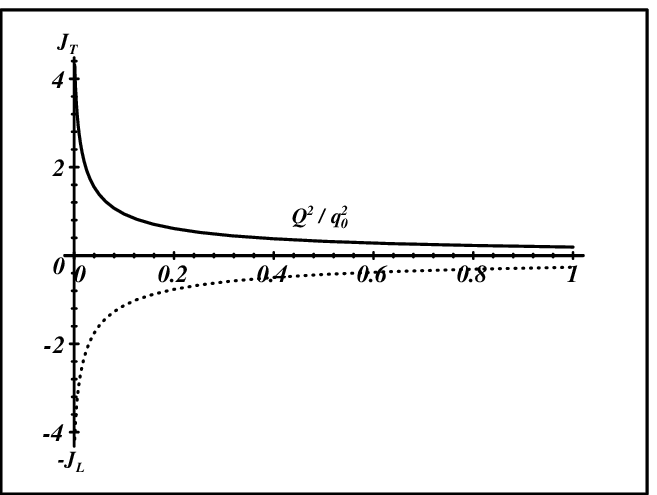}}
\vskip 3mm
      \caption{\footnotesize{Transverse and longitudinal
        contributions as a function of $Q^2/q_o^2$, for $3$ colors and
        $3$ light flavors. The value taken for the coupling constant
        is $g=0.44$.}\label{fig:plotQ}}
\end{figure}

The quantities $J_{_{T,L}}$ are coefficients quantifying the
respective contributions of the transverse and longitudinal exchanged
gluons. If we plot these quantities as a function of the parameter
$Q^2/q_o^2$, we obtain the curves of figure \ref{fig:plotQ}.  One can
see that these coefficients are decreasing very fast if $Q^2$
increases. This is of course due to the fact that the invariant mass
$Q^2$ attenuates the effect of the collinear singularities. What
happens around $Q^2/q_o^2\sim 1$ will be presented in detail by H.
Zaraket elsewhere in these proceedings \cite{Zarak1}.

Another feature of the result (\ref{eq:im-pi}) is that it is totally
free of any infrared divergence when the gluon momentum is going to
zero, while one could have expected problems to occur with the
transverse gluons (see the problems encountered in the calculation of
the damping rate of a moving quark, for instance). If we look formally
at the limit $M_\infty\to 0$ of the quantities $J_{_{T,L}}$ when
$Q^2=0$, we find
\begin{eqnarray}
&&\lim_{M_\infty\ll m_{\rm g}} J_{_{L}}\sim \ln(m_{\rm g}/M_\infty)\\
&&\lim_{M_\infty\ll m_{\rm g}} J_{_{T}}\sim \ln(m_{\rm g}/M_\infty)^2\; .
\label{eq:limits}
\end{eqnarray}
This result is interpreted as follows: the common power of the
logarithm is due to the collinear singularity (common to both the
transverse and longitudinal gluon contributions) which reappears when
its regulator $M_\infty$ is suppressed. The reason why only a
logarithm appears in this limit despite the fact that the singularity
has been described earlier as quasi linear is precisely related to the
fact that we don't have exactly a double pole but two very close
poles: the small separation between the two poles, which is
responsible for extra powers of $1/g$, does not go to zero when
$M_\infty\to 0$. In the transverse gluon contribution, there is an
extra power of this logarithm which is interpreted as a remnant of an
infrared singularity. To support this interpretation, we can introduce
by hand a magnetic mass $m_{\rm mag}$ in the effective gluon
propagator, and look at
\begin{equation}
\lim_{M_\infty\ll m_{\rm mag} 
\ll m_{\rm g}} J_{_{T}}\sim \ln(m_{\rm g}/M_\infty)\ln(m_{\rm g}/m_{\rm mag})
\; .
\end{equation}
This limit indicates the infrared nature of the extra logarithm in the
transverse gluon contribution. Equation (\ref{eq:limits}) then tells
us that the relevant regulator of this potential infrared singularity
is the {\sl quark} thermal mass when $M_\infty\gg m_{\rm mag}$, which
can be understood in terms of phase space constraints. Indeed, if
$L\to 0$ while $M_\infty > 0$, the delta functions
$\delta(R^2-M^2_\infty)$ and $\delta((P+L)^2-M^2_\infty)$ become
incompatible with bremsstrahlung, which means that $M_\infty$ prevents
the infrared singularity from having a support.

\subsection{Hard photon production}
The calculation of 2-loop contributions to hard real photon production
follows closely what has been done before for soft photons, at least
for $Q^2=0$, a condition to which I will adhere in the following.
Indeed, most of the aspects of the calculations were based on the fact
that $Q^2\ll q_o^2$. Simply, since $\bbox{q}$ cannot be neglected in
front of $\bbox{p}$, one must treat more carefully the angles between
the various 3-vectors of the problem. If we still denote by $\theta$
the angle between $\bbox{p}$ and $\bbox{q}$, the angle $\theta''$
between $\bbox{q}$ and $\bbox{l}$ is now given by
\begin{equation}
\cos\theta''={{p}\over{r}}(\cos\theta\cos\theta'+
\sin\theta\sin\theta'\cos\phi)+{{q}\over{r}}\cos\theta'\; ,
\end{equation}
where $\theta'$ and $\phi$ are defined in the same way as before.
Then, it is easy to see that (\ref{eq:R2}) is unchanged (except for
the fact that $M^2_{\rm eff}$ becomes simply $M^2_\infty$ when
$Q^2=0$), while (\ref{eq:PL2}) becomes
\begin{equation}
  \int\limits_{0}^{2\pi}\!\!\!\!{{d\phi}\over{(P+L)^2-M^2_\infty}}\approx
  {{2\pi(p+q)(2qp^2)^{-1}}\over{\left[
        \left(u+{{M^2_\infty}\over{2p^2}}+{{L^2}\over{2p^2}}\right)^2
        -{{L^2}\over{p^2}}{{M^2_\infty}\over{p^2}}\right]^{1/2}}}\; .
\label{eq:PL2hard}
\end{equation}
Again, the only dominant term in the matrix element is given by
(\ref{eq:dominant}), which makes easy the evaluation of the
bremsstrahlung contribution to hard real photons. We now obtain
\begin{eqnarray}
 && {\rm Im}\,\Pi^{^{AR}}{}_\mu{}^\mu(q_o,\bbox{q})\approx -{{e^2 g^2 N
      C_{_{F}}}\over{\pi^4}}
{T\over{q_o^2}}
(J_{_{T}}+J_{_{L}})\nonumber\\
&&\times
\int\limits_{0}^{+\infty}dp\,\left[{{p^2+(p+q_o)^2}}\right]
{{[n_{_{F}}(p)-n_{_{F}}(p+q_o)]}}\; ,
  \label{eq:hardresult}
\end{eqnarray}
with now
\begin{eqnarray}
&&J_{_{T,L}}=\int\limits_{0}^{1}{{dx}\over{x}}\,
  |\widetilde{I}_{_{T,L}}(x)|\nonumber\\
&&\qquad\times
\int\limits_{0}^{+\infty}dw\,
  {{\sqrt{w/{w+4}}\,{\rm
        tanh}^{-1}\sqrt{w/{w+4}}}\over{(w+\widetilde{R}_{_{T,L}}(x))^2
      +(\widetilde{I}_{_{T,L}}(x))^2}}\; .
\end{eqnarray}
As one can see, the structure of the bremsstrahlung contribution to
the production of hard real photons is quite similar to the case of
soft photons.  In the above expressions, the coefficients $J_{_{T,L}}$
have been defined to match exactly those which have been defined
previously, when $Q^2=0$. The only difference is in the integration
over $p$ which is now more complicated because we cannot neglect
$q_o$.  The $q_o$ dependence that comes from this integral is non
trivial now, but some limits are simple.  In the limit $q_o\ll T$, we
recover of course the result given for the production of soft real
photons. A simple result can also be obtained in the opposite limit
$q_o\gg T$, for which a simple expansion can be given for the integral
over $p$, which leads us to
\begin{equation}
  {\rm Im}\,\Pi^{^{AR}}{}_\mu{}^\mu(q_o,\bbox{q})\approx -{{e^2 g^2 \ln(2) N
      C_{_{F}}}\over{\pi^4}}
{T^2}
(J_{_{T}}+J_{_{L}})\; .
  \label{eq:hardresult-asympt}
\end{equation}

As said before, the contribution of region II is negligeable only when
$q_o$ is soft because of a small phase space, so that we must now
take it into account.  Its evaluation in the asymptotic region $q_o\gg
T$ is very easy.  Since $p_o=-\omega_{\bbox{p}}$ in this region, we
have instead of (\ref{eq:R2})
\begin{equation}
R^2-M^2_\infty\approx -2pq\left[
v+{{M^2_\infty}\over{2p^2}}\right]\; ,
\end{equation}
where I denote $v\equiv 1+\cos\theta$.  We can also check that
(\ref{eq:PL2hard}) remains valid, except for the changes $u\to v$ and
$p+q\to q-p$. The enhancement due to collinear singularities in the
terms that contains these two denominators is unchanged near $v\sim
g^2$.  Concerning the boundaries of the integration domain, we can
note that the statistical weight $n_{_{F}}(p_o)-n_{_{F}}(p_o+q_o)$ is
equal to $1$ for $-q_o\le p_o\le 0$ and equals $0$ everywhere else
(excepted in a small region of width $T$ where the transition from $1$
to $0$ takes place, but this provides corrections of relative order
$T/q_o\ll 1$).  For the $l$ and $x$ variables, the limits are $-1\le x\le
1$ and $0\le l\le l^*$ where the upper bound $l^*$ is always hard so
that it can be taken to infinity without changing the result.
Therefore, the integration over $x$ and $l$ gives again the same
factor $J_{_{T}}+J_{_{L}}$ as in (\ref{eq:hardresult-asympt}). Since
the integration over $p$ is replaced now by
\begin{equation}
\int\limits_{0}^{q_o}dp(p^2+(q_o-p)^2)={{2q_o^3}\over{3}}\; ,
\end{equation}
the contribution of region II for $q_o\gg T$ is
\begin{equation}
 {\rm Im}\,\Pi^{^{AR}}{}_\mu{}^\mu(q_o,\bbox{q})\approx -{{e^2 g^2  N
      C_{_{F}}}\over{3\pi^4}}{q_o T}
(J_{_{T}}+J_{_{L}})\; .
\label{eq:hardresult-II}
\end{equation}
It appears that the process of this region, behaving as $q_oT$ when
$q_o\gg T$, instead of $T^2$ for bremsstrahlung, is the dominant one
for the production of extremely hard photons.

\section{Comparison with existing results}

\subsection{Other thermal field theory approaches}
The production rate of soft real photons has already been evaluated at
1-loop in thermal field theory by \cite{BaierPS1,AurenBP1}. The
typical order of magnitude they found for the imaginary part of the
photon polarization tensor is $e^2g^4 \ln(1/g) T^3/q_o$, which means
that the bremsstrahlung contribution found in (\ref{eq:im-pi}) is
larger by a factor $1/g^2$.  Bremsstrahlung seems therefore to be the
dominant process for the production of soft quasi-real photons.  This
extra $1/g^2$ factor can be traced back into the strong collinear
singularity exhibited by bremsstrahlung processes. This effect can be
seen as a breakdown of the effective expansion based on the
resummation of hard thermal loops. Indeed, the powers of $1/g$
generated by these collinear divergences can change the order in $g$
of a contribution, so that the usual connection between the number of
loops and the order in $g$ is lost.

\begin{figure}[htbp]
  \centerline{\epsfxsize=7cm\epsffile{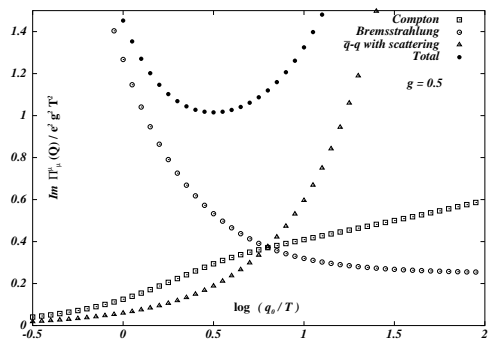}}
\vskip 3mm
      \caption{\footnotesize{Comparison of various 
          contributions to ${\rm Im}\,\Pi^\mu{}_\mu(Q)$ for a hard
          real photon.}\label{fig:comp-hard}}
\end{figure}

Concerning hard real photons, \cite{BaierNNR1,KapusLS1} calculated in thermal
field theory the contribution of processes involving a time-like
gluon, like Compton processes. Their calculation involves both 1-loop
and 2-loop diagrams, and leads to
\begin{equation}
  {\rm Im}\,\Pi^{^{AR}}{}_\mu{}^\mu(q_o,\bbox{q})\approx -{{e^2 g^2  N
      C_{_{F}}}\over{16 \pi}}
{T^2}\ln\left(
{{cq_o}\over{\alpha_{_{S}}T}}
\right)\; ,
  \label{eq:hardresult-asympt-baier}
\end{equation}
for $q_o\gg T$, where $c$ is a numerical constant equal to $0.23$. We
are now in a position to compare the contributions given by
bremsstrahlung, the $q\bar{q}$ annihilation appearing in region II,
and the Compton processes calculated by Baier et al.  For this
comparison, we take $N=3$ colors and $N_{_{F}}=2$ light flavors, and a
coupling constant $g=0.5$.\footnote{In the perturbative regime where
  $g\to 0$ and for photon energies around $T$, Compton processes are
  actually dominant because of an extra factor of order $\ln(1/g)$.} At
$Q^2=0$, the quantities $J_{_{T,L}}$ depend only on $N$ and
$N_{_{F}}$, and are $J_{_{T}}\approx 4.45$ and $J_{_{L}}\approx 4.26$
with our choice of parameters.  The evolution with the photon energy
$q_o$ of these three contributions has been plotted on figure
\ref{fig:comp-hard}.
We see that bremsstrahlung is dominating for the smaller values of
$q_o$, while the region II becomes dominant for hard enough $q_o$. For
intermediate energies, the three contributions are comparable. The sum
of the three contributions is significantly above the Compton
contribution considered alone.

\subsection{Connection with plasmon frequency calculations}
In his talk \cite{Flech1}, Flechsig presented results about the QCD
plasmon frequency near the light-cone. This calculation is also
plagued by very strong collinear singularities, very similar to those
encountered in the real photon rate. A first sight, there is however a
difference because the collinear singularities in the plasmon
frequency arise already at one loop, and come from $ggg$ and $gggg$
effective vertices.

In order to make the connection between the present work and
Flechsig's approach, it is useful to note that at the level of the
topologies, we have the following identity
\setbox1=\hbox to
8.1cm{\epsfxsize=8cm\epsffile{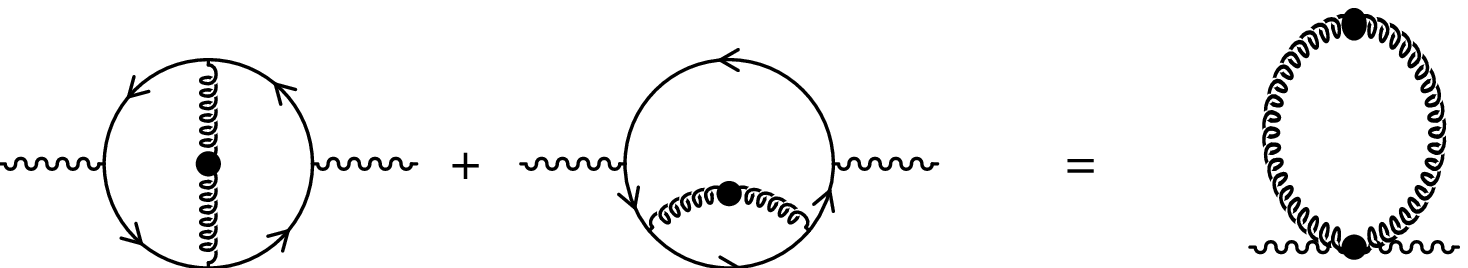}\hfill}
\begin{equation}
\raise -1cm\box1
\label{eq:tadpole}
\end{equation}
between our diagrams and some 1-loop tadpole diagram. Indeed, the
tadpole contains both the self-energy and the vertex topologies since
there are several terms contributing to the $\gamma\gamma g g$
effective vertex, obtained by exchanging a photon and a gluon. Of
course, the tadpole vanishes when one takes the trace over photon
Lorentz indices, so that this equation does not reflect an exact
algebraic identity between both sides (one would need to calculate the
tadpole beyond the HTL approximation in order to be able to identify
the two sides).  Despite this limitation, one can note that both sides
have the same denominators and therefore the same singularities.  As a
consequence, we can interpret our singularities as being collinear
singularities encountered in hard thermal vertices when some external
legs are on the light-cone\cite{FlechR1} and conclude that the
singularities encountered in our calculation are basically the same as
those encountered by Flechsig.  The only difference is that the HTL
approximation makes the tadpole vanish when one traces it over the
photon Lorentz indices, so that we have to go to two loops.

This remark suggests also a defect of the HTL approximation. Indeed,
in the calculation of the tadpole of Eq.~(\ref{eq:tadpole}) the HTL
approximation gives a zero coefficient to otherwise extremely singular
(and large after regularization) terms. Therefore, it seems that the
HTL approximation is not a good one in the presence of these strong
collinear singularities, because one would need to go beyond this
approximation to retain the relevant dominant terms in the tadpole.

\subsection{Semi-classical approach}
Photon production by a plasma has also been studied by semi-classical
methods in \cite{CleymGR2,CleymGR3,BaierDMPS1,BaierDMPS2}. Basically,
in this approach, one studies the radiative energy loss of a fast quark
undergoing multiple scatterings in a dense medium. The scattering
centers are assumed to be static, which implies that the interaction
is purely electric. This amounts to take only the longitudinal
exchanged gluons into account. As a consequence, this interaction can
be regularized by a Debye mass, and one does not have to worry about
infrared divergences.  In order to exhibit the analogy between this
semi-classical approach and the thermal field theory, it is possible
to write the bremsstrahlung contribution as given by (\ref{eq:rate})
in thermal field theory in the following form \setbox1=\hbox to
5cm{\mbox{\epsfxsize=5cm\epsffile{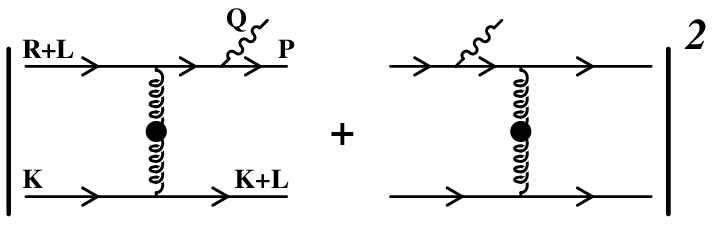}}}
\begin{eqnarray}
  &&\left.{{dN}\over{dtd{\bbox x}}}\right|_{\rm bremss}={{d{\bbox
        q}}\over{(2\pi)^32q_o}}\int{{d^4P}\over{(2\pi)^4}}
  \int{{d^4K}\over{(2\pi)^4}}
  \int{{d^4L}\over{(2\pi)^4}}\nonumber\\
  &&\quad\times\raise -7mm\box1\nonumber\\
  &&\quad\times n_{_{F}}(r_o+l_o)n_{_{F}}(k_o)
  [1-n_{_{F}}(p_o)][1-n_{_{F}}(k_o+l_o)]\nonumber\\
&&\quad\times2\pi\delta(P^2-M^2_\infty)\,
  2\pi\delta((R+L)^2-M^2_\infty)\nonumber\\
  &&\quad\times2\pi\delta(K^2-M^2_\infty)\,
  2\pi\delta((K+L)^2-M^2_\infty)
  \; ,
  \label{eq:lastform}
\end{eqnarray}
where $R\equiv P+Q$. This kind of formula looks like the starting
point of the semi-classical approach, but it differs from it by some
essential aspects. The most important one is that it includes the
contributions of the transverse exchanged gluons, which we have shown
to be quite important since $J_{_{T}}$ is as large as $J_{_{L}}$.
Neglecting it does not appear to be a good approximation. Moreover,
this formula differs from the semi-classical ones by the details of
how the Debye screening is incorporated in the gluon propagator.
Indeed, the semi-classical approach just includes some constant Debye
mass as a regulator, while in thermal field theory one takes into
account all the momentum dependence of the gluon self-energy via the
HTL resummation. Although this does not change significantly the
physics, it may affect the numerical value of the production rate.

On the other hand, these simplifications allow the semi-classical
method to handle more easily the multiple scatterings, and to show
that some new collective phenomenon may appear. More precisely,
multiple coherent scatterings may conspire in order to give a smaller
photon production rate, a phenomenon known as the
Landau-Pomeranchuck-Migdal effect \cite{LandaP1,Migda1}. In order to
extract this effect from thermal field theory, one should look at
diagrams with two and more exchanged gluons.

\section{Conclusions and perspectives}
In these proceedings, I have shown that 2-loop processes involving the
exchange of a space-like gluon play a dominant role in the real photon
production by a plasma. In particular, for the production of soft real
photons, bremsstrahlung dominates over 1-loop contributions by a
$1/g^2$ factor. Therefore, this study reconciliates thermal field
theory with the semi-classical approach.

From a more formal point of view, this phenomenon is due to very
strong collinear singularities in these processes. Although
regularized by a thermal mass of order $gT$, an enhancement by a
$1/g^2$ factor appears in the result as a remnant of these potential
divergences. The possibility of such an enhancement can make certain
contributions much larger than what one would have expected on the
basis of their number of loops, therefore resulting in some kind of
breakdown of the relationship between the order of magnitude and the
number of loops. It would be very interesting to know if even stronger
singularities can appear in higher order diagrams.

There are also other motivations to look at higher order topologies.
Indeed, the LPM effect found in the semi-classical approach seems to
indicate that equally important contributions come from the diagrams
corresponding to photon emission induced by multiple scatterings.
Moreover, the study of the effect of the thermal width made by
\cite{Niega5,Niega6} seems also to indicate contributions from higher
order diagrams.

\end{document}